\documentstyle[preprint,aps,graphicx]{revtex}
\begin{document}
\draft
\title{On convergence of the HFF expansion for 
meson production in NN collisions
} 
\author{E.Gedalin\thanks{gedal@bgumail.bgu.ac.il},
 A.Moalem\thanks{moalem@bgumail.bgu.ac.il}
 and L.Razdolskaya\thanks{ljuba@bgumail.bgu.ac.il}}
\address
{Department of Physics, Ben Gurion University, 84105
Beer Sheva, Israel}
\maketitle
\begin{abstract}
We consider the application of heavy fermion formalism based $\chi$PT
to meson production in nucleon-nucleon collisions.
It is shown that to each lower chiral order irreducible diagram
there corresponds an infinite sequence of loop diagrams which are of 
the same momentum power order. This destroys the one-to-one correspondence 
between the loop and small momentum expansion and thus rules out the
application of any finite order HFF $\chi$PT to the 
$NN\rightarrow NN \pi$ reactions. 

\end{abstract}
\ \\


There have been several attempts recently, to apply the heavy fermion 
formalism (HHF) based chiral perturbation theory 
($\chi$PT)\cite{park93,bernard95}, to calculate meson production rate 
in nucleon-nucleon 
collisions\cite{park96,cohen96,sato97,gedalin98,sato99,rocha99}. 
It is well known that in 
a fully relativistic $\chi$PT there is no one-to-one correspondence
between the loop and small momentum expansion. Such a correspondence is 
believed to be restored in the extremely non-relativistic approach of the
HFF\cite{park93,bernard95}. An essential and most important
clue to assess the validity of these calculations resides on how
rapid the HFF expansion converges. Detailed $\chi$PT calculations which
account for all contributions from tree and one loop diagrams to chiral
order D=2, show that within the frame work of the HFF, one loop
contributions are sizably bigger than the lowest-order impulse and
rescattering terms, indicating that the HFF power series expansion
does not converge fast enough and therefore may not be suitable to
calculate pion production rate in NN collisions\cite{gedalin98}. 
More recently Bernard et al.\cite{bernard98} 
and Gedalin et al.\cite{gedalin981} have shown that the HFF power series 
expansion of the nucleon propagator is on the border 
of its convergence circle. Consequently, a finite order HFF can not 
possibly predict nucleon pole terms correctly and should not be applied
to meson production. It is the purpose of the present comment to call
attention to the fact that within the framework of  the HFF $\chi$PT,  
the one-to-one correspondence between the loop and small momentum
expansion is badly destroyed for processes of sufficiently large momentum
transfer. Particularly, for each low chiral order $D$ diagram
there corresponds an infinite sequence of $n$-loop diagrams, $n=1,2,...$, 
of chiral order $D_n = D + 2n$, which have the same low momentum power
as the original diagram. Therefore, any finite chiral order HFF based
$\chi$PT calculations, can not possibly explain meson production in
NN collisions.
 
To be specific we consider pion production via the $NN \to NN \pi$
reaction. Such a process necessarily involves large
momentum transfer. The characteristic four momentum transferred at 
threshold is $Q \approx (-m_{\pi}/2, \sqrt {M m_{\pi}})$, where $M$ 
and $m_{\pi}$ are masses of the nucleon and meson produced. This stands
in marked difference with the Weinberg's standard power
counting\cite{weinberg79}, where presumably the momentum transferred
is considerably smaller of the order $Q^2 \approx m^2_{\pi}$. 
Clearly, one can not use directly the original power counting scheme
of Weinberg\cite{weinberg79}. However, we may apply 
the modified Weinberg's power counting,  a scheme tailored to deal
specifically with the production process \cite{cohen96}.
This scheme includes the following rules :\\
(i)  a $\pi NN$ vertex of zero chiral order D=0, $V^{(0)}_{\pi NN}$,
contributes a factor $ Q/F$, \\
(ii)  a pion propagator contributes a factor $Q^{-2}$,\\
(iii)  a nucleon propagator $ (v \cdot Q)^{-1} \approx m_\pi^{-1}$,\\
(iv)  a $\pi NN$ vertex of chiral order D=1, $V^{(1)}_{\pi NN}$
contributes a factor $ (k^0 Q / F M) \approx m_\pi^{(3/2)}/(F M^{1/2})$,\\
(v)  a $2\pi$NN D=1 vertex, $V^{(1)}_{\pi \pi NN}$, contributes a
factor $ k^0 Q^0/(F^2 M)$. \\
In our notations (see Fig. 1) we refer to the four-momentum squared $ Q^2
= (p_1-p_2)^2 = (v Q)^2 - {\vec Q}^2 \approx - m_\pi M$, with $v$ 
, $|{\vec Q}| \sim \sqrt {m_\pi M}, Q^0 = vQ \sim m_\pi$, $ k^0 $,  being
the nucleon four-velocity,  the transferred three-momentum, the transferred 
energy and the pion total energy,respectively. The radiative pion decay
constant 
is denoted by F. In terms of $ Q= \sqrt {m_\pi M}$ and $vQ = m_\pi$, the rules 
listed above are exactly the ones quoted by Cohen et al.\cite{cohen96}.
To calculate loop contributions one has to add three more rules :\\     
(vi) a loop integration contributes a factor $(Q^2/ 4\pi)^2$, \\ 
(vii) a  four pion vertex of zero order, $V^{(0)}_{\pi \pi \pi \pi}$,
contributes a factor $Q^2 /F^2 \sim (m_\pi M)/ F^2$, \\
(viii) a 3$\pi$NN zero order vertex, $V^{(0)}_{\pi \pi \pi NN}$,
contributes a factor $Q/F^3 \sim \sqrt{m_\pi M} / F^3$.\\
The last two factors originate, respectively,  
from the terms $\pi^2 (\partial_\mu \pi)^2/6F^2$ and 
$S^{\mu} \pi^2 \partial_\mu \pi/ 6F^3$ in the lowest-order Lagrangian.
(see for example Eqn. 2 of Ref.\cite{gedalin98}).

We now turn to demonstrate that for each low chiral order D diagram there
exist infinite sequence of loop diagrams of higher chiral order, which
have the same low momentum power as the original diagram. Consider for
example the diagrams shown in Fig. 1.
The  simplest irreducible diagram is that of the so called impulse term,
(graph 1a), corresponding to a chiral order D=1.
As shown in Ref.\cite{cohen96}, by using the rules quoted above, the low
momentum power order of this term is, 
$\Theta _0 \sim F^{-3}(m_\pi/M)^{1/2}$.  
Next, by adding two zero chiral order 
$\pi NN $ vertices, two lowest
order nucleon propagators  and one meson propagator to the diagram 1a, one
obtains the irreducible one loop diagram 1b.
We recall that a zero chiral order $\pi NN$  vertex is
proportional to the meson three momentum, i.e., 
\begin{equation}
	V_{\pi NN} = \frac{g_A}{F}S\cdot Q \tau, 
\end{equation}
where S is the nucleon spin-operator and contributes a factor 
$QF^{-1}$ (rule (i) above). Thus the two added nucleon vertices give a
factor $Q^2F^{-2}$. Likewise, a meson propagator contributes a factor
$Q^{-2}$,  the two nucleon propagators give $(v\cdot Q)^{-2}$  and the 
loop integral contributes a factor of $Q^4(4\pi)^{-2}$. Altogether the
diagram 1b has an additional factor $Q^4(4\pi F v\cdot Q)^{-2}$ with
respect to original diagram 1a. The power factor of diagram 1b is
therefore,
\begin{equation}
	\Theta_3 = \Theta_1 \frac{Q^4}{(4\pi F)^2 (v\cdot Q)^2}.
\end{equation}
With  $4\pi F \sim M, v\cdot Q \sim m_\pi$,  $\Theta_3 = \Theta_1$, so
that 
although higher in chiral order, the diagram 1b is of the same order as
the diagram 1a. Similarly, by adding progressively, two zero order $\pi NN$
vertices, a pion propagator and two nucleon propagators, as mentioned
above, one obtains the other irreducible n-loop diagrams shown in Fig. 1.
  
By making use of the same power counting rules as above, the
momentum power of n-loop diagram would be,
\begin{equation}
	\Theta_{2n+1} =
	 \Theta_1\left(\frac{Q^4}{(4\pi f)^2 (v\cdot Q)^2}\right)^n = 
	\Theta_1.
\end{equation}
Thus for the impulse diagram 1a there exists an infinite sequence of  
n-loop diagrams, $n=1,2,...$ of chiral order $2n + 1$ all having the
same characteristic momentum power as the lowest chiral order tree
diagram. Quite obviously, such a sequence of loop diagrams can be
constructed in a similar manner for any irreducible diagram that may 
contribute to the production process.

Thus the basic principle of the HFF $\chi$PT of one-to-one correspondence 
between the loop and small momentum expansion is badly destroyed. The
primary production amplitude becomes the sum over 
infinite sequences of loop diagrams all having the power order, and thus
excluding the possibility
that a finite chiral 
order HFF based $\chi$PT calculations can explain meson production in 
NN collisions. This result,  along with the observation made in
Refs.\cite{bernard98,gedalin981},
that the HFF series of the nucleon propagator 
is on the border of its convergence circle, lead us to conclude that  
the $NN\rightarrow NN \pi$ production process  falls
outside the HFF validity domain.

\vspace{1.5 cm}
{\bf Acknowledgments} This work was supported in part 
by the Israel Ministry Of Absorption.

\begin{figure}
\includegraphics[scale=0.6]{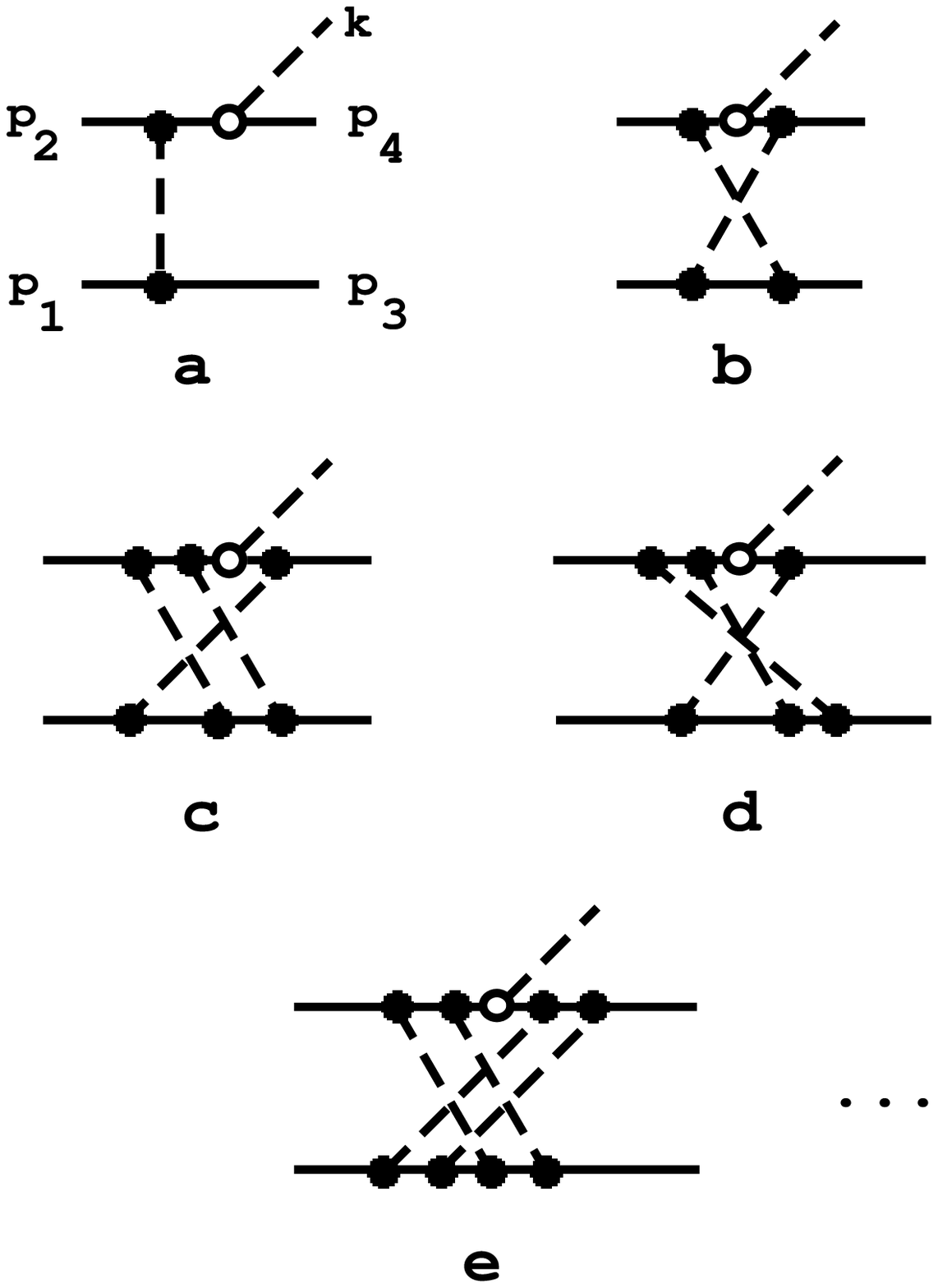} 
\caption{An infinite sequence of diagrams contributing to the
production process (see text). A solid line stands for a nucleon and a
dashed line represents a meson. The black dot are zero chiral order 
$\pi NN$ vertex, an open circle denotes a $\pi NN$ vertex of chiral order
1. 
Shown in the figure are : the irreducible
impulse diagram (1a), a one-loop diagram (1b), two-loop diagrams (1c, 1d)
and three-loop diagram (1e). The ellipsis denote all other loop diagrams 
of the sequence.
 }
\label{xfig3}
\end{figure}  

\end{document}